\documentclass[aps,prd,preprint,preprintnumbers,showpacs,showkeys,nofootinbib,superscriptaddress,floatfix]{revtex4-1}
\usepackage{amsmath,amsfonts,amssymb}
\usepackage{graphicx}
\usepackage{multirow}
\usepackage{bigstrut}
\usepackage{makecell}
\usepackage{xcolor}

\begin{document}

\title{Decay processes of a pseudoscalar $D(2900)$}
\author{Brenda B. Malabarba}
\email[E-mail: ]{brenda@if.usp.br}
\affiliation{Universidade de Sao Paulo, Instituto de Fisica, C.P. 05389-970,
Sao Paulo, Brazil.}

\author{K.\,P.\,Khemchandani}
\email[E-mail: ]{kanchan.khemchandani@unifesp.br}
\affiliation{Universidade Federal de S\~ao Paulo, C.P. 01302-907, S\~ao Paulo,
Brazil.}

\author{A.\,Mart\'inez Torres}
\email[E-mail: ]{amartine@if.usp.br}
\affiliation{Universidade de Sao Paulo, Instituto de Fisica, C.P. 05389-970,
Sao Paulo, Brazil.}

\date{\today}
\begin{abstract}
We study the decay properties of a $D(2900)$ state, with spin-parity $0^-$, whose existence was proposed in an earlier study of the $D K\bar K$ and coupled channel system. It was found in the former work that a $D$-meson  appears with a mass of about 2900 MeV from the three-body dynamics while the charmless subsystem of the  pseudoscalar mesons forms the $f_0(980)$ resonance. Motivated by the recent experimental investigations of $D$-mesons around 3000 MeV, we now study the two-body decays of $D(2900)$. We find that the nature of the said state sets the main decay channels to be $D^*\pi$, $D^{*}_s \bar K$ and $D^*_{s0}(2317)\bar K$. It turns out that  decay width to the last one is the largest, making the $D^*_{s0}(2317)\bar K$ system to be the most favorable one to look for a signal of $D(2900)$. We compare the decay properties of our state with those of the $D$-meson states, proposed within quark models, near 3000 MeV. We hope that our findings and discussions can be useful for the future experimental investigations of charm mesons around  3000 MeV.
\end{abstract}
\maketitle

\section{Introduction}
With the access to the charm physics becoming more available in recent times, it seems possible to uncover the spectra of charm hadrons with  certainty. In the last decade the BaBar and LHCb Collaborations have brought forward informations on charm mesons in the mass region above~2.4 GeV \cite{delAmoSanchez:2010vq,Aaij:2013sza,Aaij:2016fma,Aaij:2019sqk}. Though the evidence for some of the  states has been confirmed by both the Collaborations, the quantum numbers of such $D$-mesons are still under discussions. Structures at the highest mass known so far, around 3000 MeV, have been observed by the LHCb Collaboration~\cite{Aaij:2013sza,Aaij:2016fma}. In the former work, LHCb reports a signal around 3000 MeV in the $D^{*+}\pi^-$, $D^{+}\pi^-$ and $D^{0}\pi^+$ mass spectra. The structure found in the $D^{*+}\pi^-$ spectrum is found to be compatible with an unnatural parity  while  that in the $D^{+}\pi^-$ and $D^{0}\pi^+$ spectra seems compatible with a natural parity assignment.  The former one is denoted by $D_J(3000)$ and the latter by  $D^*_J(3000)$ in Ref.~\cite{Aaij:2013sza}.  Further,  a spin 2 $D$-meson has been found in the $D^+\pi^-$ amplitude in a later work~\cite{Aaij:2016fma}, and though it is labelled as $D_2^*(3000)$, the central value of its mass is  3200 MeV.  Interestingly, in a more recent study of $B^-$ decay to $D^{*+}\pi^-\pi^-$~\cite{Aaij:2019sqk}, no signal of $D_2^*(3000)$ is found in the $D^*\pi$ system. All these findings have motivated a series of  studies of the $D$-meson spectrum.

Different model calculations have been presented in Refs.~\cite{Ebert:2009ua,Sun:2013qca,Yu:2014dda,Lu:2014zua,Xiao:2014ura,Godfrey:2015dva,Song:2015fha,Batra:2015cua,Li:2017zng,Gupta:2018zlg,Badalian:2020ngz,Gandhi:2021col} to understand the properties of $D_J(3000)$, within relativistic formalisms and by  considering a variety of potentials, like, an effective interaction arising from the sum of  a one-gluon exchange term and long-range confining potentials, those based on heavy quark symmetry and chiral symmetry, etc.  There seems to be a common finding in all these works, which is that states with quantum numbers $n\,^{2s+1}\!J_L$ = $2\,^{1}\!P_1$, $2\,^{3}\!P_1$, $3\,^{1}\!S_0$, $1\,^{1}\!F_3$, $1\,^{3}\!F_3$ have a mass value of around 3000 MeV, and are all compatible with $D_J(3000)$. The decay properties of these aforementioned states, however, seem to be different. Authors of different works favor different spectroscopic assignments for $D_J(3000)$, though they cannot strongly exclude association with other possible quantum numbers since the information available from experiments is scarce  and the quality of statistics of the data is poor at this point. For example, Refs.~\cite{Xiao:2014ura,Yu:2014dda,Li:2017zng,Gandhi:2021col} suggest  attributing $2P(1^+)$ to $D_J(3000)$ and indicate systems like $D^*_2(2460) \pi$, $D^*\pi$, $D_1^*(2600) \pi$ to be important decay channels. On the other hand, the authors  of Ref.~\cite{Lu:2014zua}  estimate the decay widths of the states related to $2\,^{1}\!P_1$, $2\,^{3}\!P_1$, $3\,^{1}\!S_0$, $1\,^{1}\!F_3$, $1\,^{3}\!F_3$  quantum numbers, including decays to lighter $D$-mesons in the spectra. In these latter calculations the widths of the $2P$ and $1F$ states turn out to be larger (270-500 MeV) than in other works, which lay far from the experimental data. Thus, the authors exclude all possibilities except $3{}^1S_0$. On the basis of other arguments, Refs.~\cite{Godfrey:2015dva,Song:2015fha,Badalian:2020ngz}  suggest  $3{}^1S_0$ to be the favored quantum numbers and find significant branching ratios for the decay to  $D\rho$, $D^*\rho$, $D^*\pi$, etc. 
 
The author of Ref.~\cite{Xiao:2016kak} proposes a very different description for $D_J$, which is that it can be either a $Df_2$ or $D_1 \rho$ molecular state with $J^\pi = 2^-$, though  the two states  (found in $Df_2$ and  $D_1 \rho$ systems) have very different widths. Besides,  $Df_2$ and $D_1 \rho$  could be treated as coupled channels.

From all these studies, one should expect a rich spectrum of $D$-mesons to show up in the  $D^*\pi$ invariant mass distribution, around 3000 MeV. Though this does not seem to be the case so far,  the picture should become clearer when higher statistics data is obtained in future.

With the expectations of more experimental investigations occurring in future, to better understand the properties of charmed mesons around 3000 MeV and test the series of interesting predictions made by the works mentioned above, we find it timely to study the properties of a $J^\pi=0^-$ $D$-meson whose existence was predicted by some of the authors of the present work in Ref.~\cite{MartinezTorres:2012jr}. In this former work a $D$-meson with mass around 2900 MeV was found to arise from the three-body dynamics in the $DK\bar K$ system. In Ref.~\cite{MartinezTorres:2012jr}, the same system was studied by solving few-body equations as well as through QCD sum rules by writing  correlation functions  in terms of currents representing the $D_{s0}^*(2317)\bar K$ and $Df_0(980)$ systems. Both methods lead to the finding that a $D$-meson state, with spin-parity $0^-$, arises with a mass around 2900 MeV.  Further, a width of around 55 MeV was determined from the three-body amplitude obtained in our former work. Incidentally, the formation of a state from $Df_0(980)$ dynamics was also concluded in Ref.~\cite{Debastiani:2017vhv}, where a state with a mass around 2833 MeV, but with a  narrower width was found.  However, coupled channels like, $D\pi\pi$, $D\pi\eta$, were not considered explicitly in Ref.~\cite{Debastiani:2017vhv}, which can be the reason for finding a narrower width. We shall refer to this state as $D(2900)$ in the following discussions. 

In the present work we study the main two-body decay channels of  $D(2900)$~\cite{MartinezTorres:2012jr}, which are $D^* \pi$, $D^* \bar K$ and $D_{s0}^*(2317) \bar K$. We find that the decay width for  $D(2900) \to D_{s0}^*(2317) \bar K$ turns out to be the largest and, thus, conclude that $D_{s0}^*(2317) \bar K$ should be an ideal channel to look for a signal of $D(2900)$. We also discuss that the branching ratios of $D(2900)$ to decay channels considered as important for states predicted within quark models, like, $D^*\rho$, $D\rho$, $D^*_2(2460) \pi$, $D_1^*(2600) \pi$, etc., should be much smaller. Thus, $D(2900)$ can be  distinguished from the states predicted by the quark models discussed above.   Such  findings  should be  useful in experimental studies of $D$-mesons around 3000 MeV.

\section{Formalism}
A $D$-meson arising from hadron coupled channel  dynamics, studied within two distinct formalisms, was found in Ref.~\cite{MartinezTorres:2012jr}. We find it useful to discuss the formalisms and findings of Ref.~\cite{MartinezTorres:2012jr} briefly here, since the properties of the proposed $D(2900)$ are going to be essential  in deducing its main decay mechanisms and decay channels. One of the formalisms considered in Ref.~\cite{MartinezTorres:2012jr} consisted of solving few-body equations for the channels of three-pseudoscalar systems coupling to total charm $+1$ and strangeness zero: $D^0K^+K^-$, $D^0K^0\bar K^0$, $D^0\pi^+\pi^-$, $D^0\pi^-\pi^+$, $D^0\pi^0\pi^0$, $D^0\pi^0\eta$, $D^+K^0K^-$, $D^+\pi^-\pi^0$, $D^+\pi^-\eta$, $D^+\pi^0\pi^-$.  The input two-body amplitudes were determined by solving the Bethe-Salpeter equation with the kernels deduced from chiral and heavy quark symmetry Lagrangians. Such two body amplitudes  carry the information of the dynamical generation of $D_s(2317)$ and scalar resonances: $\sigma(600)$, $f_0(980)$ and $a_0(980)$, in the $DK$, $\pi\pi-K\bar K$ and $\pi\eta-K\bar K$subsystems, respectively.   That is, if the two body amplitudes are scanned in an  isospin configuration and energy region corresponding to the states mentioned above, they show formation of a resonance in the form of a peak on the real axis or in the form of a pole in the complex energy plane. The interaction in the remaining subsystem $D\bar K$ is also attractive. In fact, more recent investigations indicate  formation of an exotic state in the $D\bar K$ system (see Ref.~\cite{Molina:2020hde}, which is an update of Ref.~\cite{Molina:2010tx}).   The three-body amplitudes obtained with such two-body inputs, when projected on the total isospin 1/2, while keeping the isospin for the charmless subsystem to be zero, exhibited a peak at a total energy of $\sim$2900 MeV in Ref~\cite{MartinezTorres:2012jr}.  The state was found when the invariant mass of the charmless subsystem was around the mass of $f_0(980)$. The findings of Ref~\cite{MartinezTorres:2012jr} were interpreted as formation of an effective $Df_0$ moleculelike state with mass around 2900 MeV. 

The same problem was also studied within another formalism, based on QCD sum rules, in our previous work. In this case, two-point correlation functions were written in terms of interpolating molecular currents for the $Df_0$ and $D_s(2317) \bar K$ systems. A good convergence of the operator product expansion series was encountered by considering condensates up to dimension seven on the QCD side and  by applying a Borel transformation. A pole plus continuum description was considered to describe the spectral density from the phenomenological point of view. As a consequence, stable mass values were found around 2900 MeV in both cases, with the current-state coupling being two time bigger for the $Df_0$ current. The precise mass values obtained, with uncertainties, in the case of the $Df_0$ current can be summarized as $2926\pm 237$ MeV.

Both studies indicate the existence of a $D$-meson with spin-parity $0^-$ and mass around 2900 MeV, arising, dominantly, from the $Df_0$ dynamics. A width of about 55 MeV was determined for the  state, from the three-body amplitude. Though such a mass value is compatible with that of $D_J(3000)$ discovered in the LHCb data~\cite{Aaij:2013sza},  which is $2971.8\pm 8.7$ MeV, the width is smaller than the experimental value $188.1 \pm 44.8$ MeV. In any case, it is difficult to discuss any relation between the two states since very limited information is available from the experimental data. The $D_J(3000)$ state in the experimental data appears close to the upper limit of the mass spectra, and, hence,  systematic uncertainties on the properties of $D_J(3000)$ could not be determined in Ref.~\cite{Aaij:2013sza}. The results obtained in our present work should be useful in the identification of a  $D$-meson with $Df_0$  molecular nature in future experimental investigations. 

\begin{figure}[h!]
\includegraphics[width=0.9\textwidth]{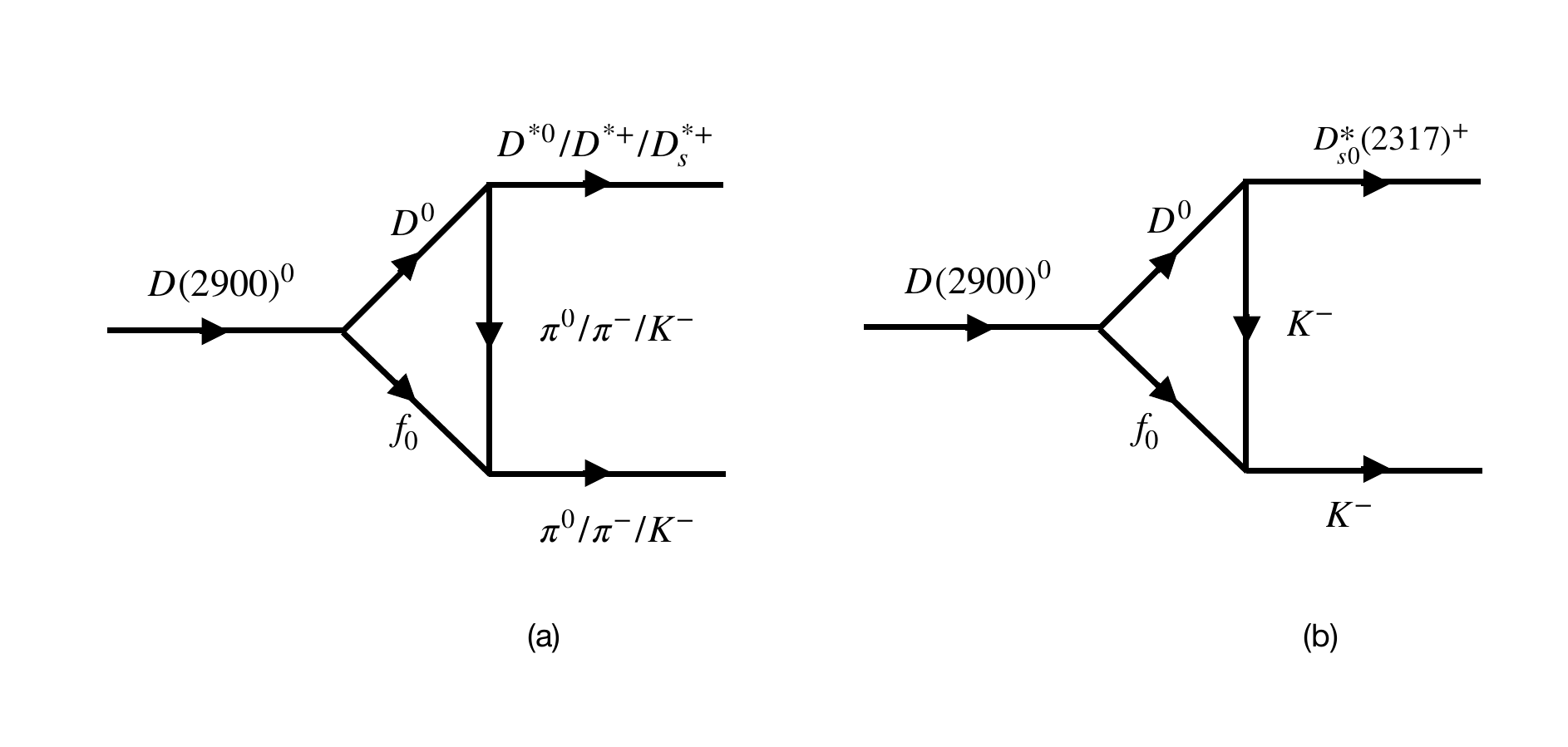}
\caption{Diagrams showing  the decay of $D(2900)$ to the different possible final states.}\label{diagram1}
\end{figure}

We are now at a position to discuss the main decay channels of $D(2900)$. Since its nature is a $Df_0$ molecular state, it must primarily disintegrate into its constituents, which can subsequently interact, leading to other decay channels through a loop. Keeping in mind that  the properties of $f_0(980)$ can be understood, essentially, by considering the contributions  from $\bar K K$ and $\pi \pi$ dynamics~\cite{pdg}, we can deduce  the decay process of $D(2900)$ to proceed through the loops shown in Fig.~\ref{diagram1}.  We can then enlist the main decay channels of the state with electric charge zero to be $D^{*0}\pi^0$, $D^{*+}\pi^-$, $D^{*+}_s K^-$ and $D^*_{s0}(2317)^+ K^-$.

We have already mentioned that  $D(2900)$ and $f_0(980)$ can be interpreted as moleculelike states. We would like to add that similar is the case of $D_{s0}^*(2317)$, which is interpreted as a $DK$ bound state within several model calculations~\cite{vanBeveren:2003kd,Barnes:2003dj,Kolomeitsev:2003ac,Szczepaniak:2003vy,Mehen:2005hc,Gamermann:2006nm,Guo:2006fu,Faessler:2007gv,Flynn:2007ki,Liu:2012zya,Cleven:2014oka,Albaladejo:2016hae}, as well as from lattice QCD analyses~\cite{Mohler:2013rwa,Torres:2014vna,Cheung:2020mql}.
In such a situation, the vertices $D(2900)^0\to D^0f_0(980)$, $D_{s0}^*(2317)^+\to D^{0}K^+$ and $f_0\to \pi^0\pi^0,~\pi^+\pi^-,~ K^+K^-$, shown in Fig.~\ref{diagram1}, can be all written in terms of their respective couplings (summarized in Table~\ref{couplings}), together with the effective fields related to each of the mesons involved in the vertex. In Table~\ref{couplings}, we provide the couplings obtained from model calculations and compare them with those extracted from the experimental data or lattice simulations, when available. It can be seen that the values coming from the model calculations are in good agreement with the information known from the experimental/lattice data.
\begin{table}[h!]
\caption{Couplings for the different vertices in Fig.~\ref{diagram1} from model calculations as well as those determined from the experimental data or lattice computations. An asterisk in the superscript indicates that the coupling was obtained following the cited model. A double asterisk implies that only  the modulus value is available. Also, note that the couplings of $f_0(980)$ to $\pi^0\pi^0$ and $\pi^+\pi^-$ are related through a factor $\sqrt{2}$ in 
Ref.~\cite{Ambrosino:2006hb}. }\label{couplings}
 \begin{tabular}{ccc}
 \hline
Vertex&Model couplings (MeV)& Experimental/lattice couplings\\
\hline\hline
$D(2900)^0\to D^0f_0(980)$& $\left(7259.63 - i 667.579\right)$~\cite{MartinezTorres:2012jr}&--\\
&&\\
$f^0\to \pi^0\pi^0$&$-\dfrac{1}{\sqrt{3}}\left(597.99 -i 2028.5\right)$~\cite{Oller:1997ti}$^*$&$\dfrac{1}{\sqrt{2}}\left(-1430\pm 10 ^{+10}_{-60} {}^{+30}_{-600}\right)$~\cite{Ambrosino:2006hb}\\
&&\\
$f_0\to \pi^+\pi^-$ &$-\dfrac{1}{\sqrt{3}}\left(597.99 -i 2028.5\right)$~\cite{Oller:1997ti}$^*$& $-1430\pm 10 ^{+10}_{-60} {}^{+30}_{-600}$~\cite{Ambrosino:2006hb}  \\
&&\\
$f_0\to K^+K^-$&$\dfrac{1}{\sqrt{2}}\left(3894.91+i1328.01\right)$~\cite{Oller:1997ti}$^*$&$3760 \pm 40^{+150}_{-80} {}^{+1160}_{-480}$~\cite{Ambrosino:2006hb} \\
&&\\
\multirow{2}{*}{$D_{s0}^*(2317)^+\to D^{0}K^{+}$}&$-\dfrac{1}{\sqrt{2}}\left(9080\pm2530\right)$~\cite{Gamermann:2006nm}&$\dfrac{1}{\sqrt{2}}\left(12600\pm1500 \right)$~\cite{Torres:2014vna}$^{**}$  \\
& (in agreement with~\cite{Guo:2006fu,Mehen:2005hc,Faessler:2007gv})&\\
\\\hline
\end{tabular}
\end{table}

The coupling of the state $D(2900)^0\to D^0f_0(980)$, given in Table~\ref{couplings}, is calculated using the method followed in Refs.~\cite{MartinezTorres:2008gy,Malabarba:2020grf}, where the two-body amplitude is assumed to be proportional to the three-body amplitude near the peak region. Following these former works, we can write $T_{Df_0}=\alpha T_{D\left[ K\bar K\right]_{I=0}}$, where $\alpha$ is a proportionality constant, which can be determined using the unitarity condition for the $Df_0$ scattering amplitude
\begin{align}
\mathcal{I}m\left\{T^{-1}_{Df_0}\right\}= \frac{|\vec p_{Df_0}|}{8\pi \sqrt{s_{Df_0}}},\label{unitary}
\end{align}
with $\vec p_{Df_0}$ being the center of mass momentum and $\sqrt{s_{Df_0}}$ is taken as the mass of $D(2900)$. Using  Eq.~(\ref{unitary}) and the three-body amplitude of Ref.~\cite{MartinezTorres:2012jr}, we can determine the relation between the effective $Df_0$  amplitude  and $T_{D\left[ K\bar K\right]_{I=0}}$. Further, assuming a Breit-Wigner form for the $Df_0$  amplitude, we can then determine the coupling $g_{Df_0}$ as
\begin{align}
g^2_{Df_0}=\alpha\, i \,M_{D(2900)}~\Gamma_{D(2900)}~T_{D\left[ K\bar K\right]_{I=0}}.
\end{align}
Using the value of the three-body amplitude, at the peak position, $T_{D\left[ K\bar K\right]_{I=0}}$, we get $g_{Df_0}=\left(7259.63 - i667.579\right)$ MeV. 

Considering now the value of $g_{Df_0}$, we can calculate the width of $D(2900)$ through
\begin{align}
\Gamma_{D(2900)}=\frac{1}{8\pi} \frac{|\vec p_{Df_0}|}{M_{D(2900)}^2} |g_{Df_0}|^2,\label{eq:width}
\end{align}
and obtain a width of the order of 55 MeV, which indeed coincides with the value determined in Ref.~\cite{MartinezTorres:2012jr}.

To calculate the  diagram in Fig.~\ref{diagram1}a, we also require the following Lagrangian for the vector-pseudoscalar-pseudoscalar (VPP) vertex:
\begin{eqnarray}\nonumber
\mathcal{L}_{VPP}&=&-ig_{VPP}\langle V_\mu \left[P,\partial_\mu P\right] \rangle\\&=& -i~\mathcal{I}_{VPP}~g_{VPP}~V_\mu\left(\phi_L\partial^\mu \phi_H-\phi_H\partial^\mu \phi_L\right),\label{vpp1}
\end{eqnarray}
where $\mathcal{I}_{VPP}$ is an isospin factor arising from the trace in the Lagrangian, $V_\mu$ is a (heavy) vector meson field, and $\phi^H$ ($\phi^L$) represents heavy (light) pseudoscalar meson field. We use the following matrices for the mesons
\begin{align} \nonumber
P =
\left( \begin{array}{cccc}
\dfrac{\pi^0}{\sqrt{2}} + \dfrac{\eta}{\sqrt{6}}+\dfrac{\eta_c}{\sqrt{12}} & \pi^+ & K^{+}&\bar D^0\\\vspace{0.2cm}
\pi^-& -\dfrac{\pi^0}{\sqrt{2}} + \dfrac{\eta}{\sqrt{6}} +\dfrac{\eta_c}{\sqrt{12}}& K^{0}&D^-\\
K^{-} &\bar{K}^{0} & \dfrac{-2\eta }{\sqrt{6}} +\dfrac{\eta_c}{\sqrt{12}}&D_s^-\\
D^0&D^+&D_s^+&\dfrac{-3\eta_c}{\sqrt{12}}
\end{array}\right),
\end{align}
\begin{align}
V^\mu =
\left( \begin{array}{cccc}
\frac{\rho^0 + \omega}{\sqrt{2}} & \rho^+ & K^{*^+}&\bar D^{*0}\\\vspace{0.2cm}
\rho^-& \frac{-\rho^0 + \omega}{\sqrt{2}} & K^{*^0}&D^{*-}\\\vspace{0.2cm}
K^{*^-} &\bar{K}^{*^0} & \phi &D^{*-}_s\\
D^{*0}&D^{*+}&D^{*+}_s&J/\psi
\end{array}\right)^\mu.
\end{align}

The coupling $g_{VPP}$, in Eq.~(\ref{vpp1}) is determined as
\begin{align}
g_{VPP}=\frac{m_\rho}{2f_\pi}\frac{m_{D^*}}{m_{K^*}} \sim 9.3,
\end{align}
where the factor $m_{D^*}/m_{K^*}$ has been included, following Ref.~\cite{Liang:2014eba}, to consider the presence of heavy mesons in the vertices.

\begin{figure}[h!]
\includegraphics[width=0.35\textwidth]{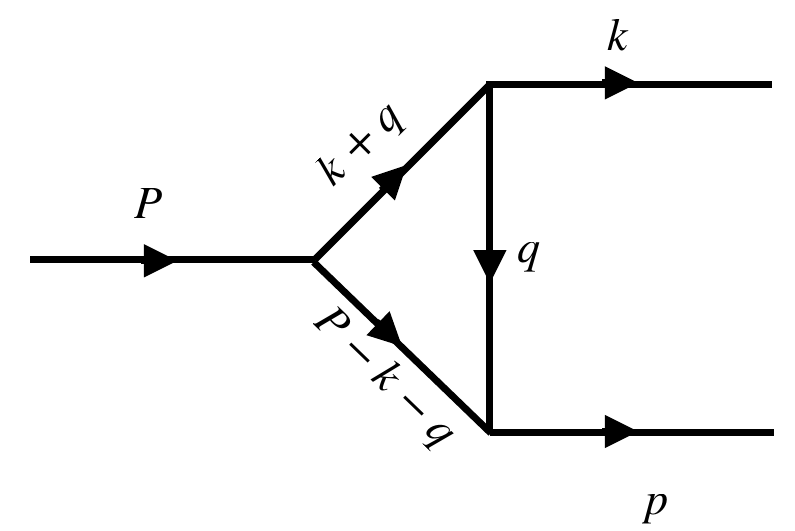}
\caption{Diagram showing momenta labels.}\label{diagram2}
\end{figure}

Using the momenta label provided in Fig.~\ref{diagram2}, 
we can write Eq.~(\ref{vpp1}) as
\begin{eqnarray}
\mathcal{L}_{VPP}& =-i~\mathcal{I}_{VPP}~g_{VPP}~\epsilon_\mu\left(-i \left[k+q\right]^\mu-iq^\mu\right)\\&=-\mathcal{I}_{VPP}~g_{VPP}~\epsilon_\mu\left(k^\mu+2q^\mu\right).
\end{eqnarray}

We can now write the amplitude for the diagram in  Fig.~\ref{diagram1}a, using the relation $i\mathcal{L}=-it$, as
\begin{eqnarray}\nonumber
it_a&=&\int\frac{d^4q}{\left(2\pi\right)^4} it_{D(2900)^0\to D^0f_0} it_{f^0\to P_1P_2}i t_{VPP}\frac{i}{\left(k+q\right)^2-m_{D}^2}\frac{i}{\left(P-k-q\right)^2-m_{f_0}^2}\frac{i}{q^2-m_{P_1}^2}\\
&=&-\int\frac{d^4q}{\left(2\pi\right)^4} \frac{g_{D(2900)^0\to D^0f_0}~  g_{f^0\to P_1P_2}~\left[ \mathcal{I}_{VPP}~g_{VPP}~\epsilon_\mu\left(k\right)\left(k^\mu+2q^\mu\right)\right] }{\left[\left(k+q\right)^2-m_{D}^2\right]\left[\left(P-k-q\right)^2-m_{f_0}^2\right]\left[q^2-m_{P_1}^2\right]}\label{lagr_a},
\end{eqnarray}
where $m_{P_1}$ is the mass of the pseudoscalar meson with the momentum $q$.
Using the Lorenz condition, the amplitude for the process becomes, 
\begin{eqnarray}
t_a&=&
2i ~g_{D(2900)^0\to D^0f_0}~  g_{f^0\to P_1P_2}~ \mathcal{I}_{VPP}~g_{VPP}~\epsilon_\mu \left(k\right) \Biggl\{\int\frac{d^4q}{\left(2\pi\right)^4} \frac{q^\mu}{\left[\left(k+q\right)^2-m_{D}^2\right]} \Biggr.\nonumber\\
&\times&  \frac{1}{\left[\left(P-k-q\right)^2-m_{f_0}^2\right]\left[q^2-m_{P_1}^2\right]}\Biggl.\Biggr\},\label{tensoreq1}
\end{eqnarray}
with the values of  $\mathcal{I}_{VPP}$ given in Table~\ref{tableiso}.  Further, following  the Passarino-Veltman reduction for tensor integrals, we can write
\begin{eqnarray}
t_a&=&
2i ~g_{D(2900)^0\to D^0f_0}~  g_{f^0\to P_1P_2}~ \mathcal{I}_{VPP}~g_{VPP}~\epsilon_\mu \left(k\right) \biggl\{a~k^\mu+ b~P^\mu\biggr\},\label{tensoreq2}
\end{eqnarray}
out of which, only the second term survives, once again, due to the Lorenz condition. Hence, we do not need to find the coefficient $a$ but we need to determine $b$. For this, let us call the integral in Eq.~(\ref{tensoreq1}) [which is equal to the terms in the curly bracket in Eq.~(\ref{tensoreq2})] as $I^\mu$.  Then,  we can get a set of equations by contracting the integral with the different four vectors 
\begin{align}\nonumber
k\cdot I&=a~k^2+b~k\cdot P\\
P\cdot I&=a~P\cdot k+b P^2,
\end{align}
which leads to
\begin{align}
b=\frac{P\cdot k~k\cdot I-k^2~P\cdot I}{\left(k\cdot P\right)^2-k^2 P^2},\label{defnb}
\end{align}
where
\begin{align}\nonumber
k\cdot I&=\int\frac{d^4q}{\left(2\pi\right)^4}\frac{k\cdot q}{\left[\left(k+q\right)^2-m_{D}^2\right]\left[\left(P-k-q\right)^2-m_{f_0}^2\right]\left[q^2-m_{P_1}^2\right]},\\
P\cdot I&=\int\frac{d^4q}{\left(2\pi\right)^4}\frac{P\cdot q}{\left[\left(k+q\right)^2-m_{D}^2\right]\left[\left(P-k-q\right)^2-m_{f_0}^2\right]\left[q^2-m_{P_1}^2\right]}.
\end{align}

\begin{table}[b!]
\caption{The values of the isospin factor, $\mathcal{I}_{VPP}$, obtained by calculating the trace in Eq.~(\ref{vpp1}), for the different vector-pseudoscalar-pseudoscalar vertices shown in Fig.~\ref{diagram1}a. }\label{tableiso}
 \begin{tabular}{cc}
 \hline
Vertex&$\mathcal{I}_{VPP}$\\
\hline\hline
$D^0D^{*0}\pi^0$&$-1/\sqrt{2}$ \\
$D^0D^{*+}\pi^-$&$-1$ \\
$D^0D_s^{*+}K^-$&$-1$ \\
\hline
\end{tabular}
\end{table}

Writing the previous equations explicitly in the center of mass frame, we have
\begin{align}\label{I1}
k\cdot I&=\int\frac{d^4q}{\left(2\pi\right)^4}\frac{k^0q^0-\vec k\cdot \vec q}{\left[\left(k+q\right)^2-m_{D}^2\right]\left[\left(P-k-q\right)^2-m_{f_0}^2\right]\left[q^2-m_{P_1}^2\right]},\\
P\cdot I&=\int\frac{d^4q}{\left(2\pi\right)^4}\frac{P^0 q^0}{\left[\left(k+q\right)^2-m_{D}^2\right]\left[\left(P-k-q\right)^2-m_{f_0}^2\right]\left[q^2-m_{P_1}^2\right]}.\label{I2}
\end{align}
To determine Eq.~(\ref{defnb}), we need to solve integrals on terms proportional to $\left(q^0\right)^0$ and to $\left(q^0\right)^1$. We can integrate Eqs.~(\ref{I1}) and (\ref{I2}) on $q^0$ analytically, through  Cauchy's theorem. To do this we rewrite Eqs.~(\ref{I1}) and (\ref{I2}) to exhibit the $q^0$ dependence
\begin{align}\nonumber
k\cdot I&=k^0\int\frac{d^3q}{\left(2\pi\right)^3}\int\frac{dq^0}{\left(2\pi\right)}\frac{q^0}{\left[\left(k^0+q^0\right)^2-\omega_{D}^2\right]\left[\left(P^0-k^0-q^0\right)^2-\omega_{f_0}^2\right]\left[\left(q^{0}\right)^2-\omega_{P_1}^2\right]}\\\nonumber
&-\int\frac{d^3q}{\left(2\pi\right)^3}\left(\vec k\cdot \vec q\right)\int\frac{dq^0}{\left(2\pi\right)}\frac{1}{\left[\left(k^0+q^0\right)^2-\omega_{D}^2\right]\left[\left(P^0-k^0-q^0\right)^2-\omega_{f_0}^2\right]\left[\left(q^{0}\right)^2-\omega_{P_1}^2\right]},\\
P\cdot I&=P^0\int\frac{d^3q}{\left(2\pi\right)^3}\int\frac{dq^0}{\left(2\pi\right)}\frac{ q^0}{\left[\left(k^0+q^0\right)^2-\omega_{D}^2\right]\left[\left(P^0-k^0-q^0\right)^2-\omega_{f_0}^2\right]\left[\left(q^{0}\right)^2-\omega_{P_1}^2\right]},
\end{align}
where
\begin{align}\nonumber
\omega_{D}&=\sqrt{\left(\vec k+\vec q\right)^2+m_D^2},\nonumber\\
\omega_{f_0}&=\sqrt{\left(\vec k+\vec q\right)^2+m_{f_0}^2},\nonumber\\
\omega_{P_1}&=\sqrt{\vec{q~}^2+m_{P_1}^2}.
\end{align}
Let us denote the integrand proportional to $\left(q^0\right)^0$ by $\mathcal{I}^0$ and the one proportional to  $(q^0)^1$ by $\mathcal{I}^1$. Closing the contour clockwise in the complex plane, we get
\begin{align}
\int\frac{dq^0}{\left(2\pi\right)}\mathcal{I}^n=-i\frac{\mathcal{N}_n}{\mathcal{D}},
\end{align}
where
\begin{align}\nonumber
\mathcal{N}_0&=2P^0k^0\omega_{f_0}\omega_{P_1}-\left(P^0\right)^2\omega_{f_0}\left(\omega_{P_1}+\omega_D\right)+\left(\omega_{f_0}+\omega_D\right)\biggl[\left(\omega_{f_0}+\omega_{P_1}\right)\left(\omega_{P_1}+\omega_D\right)\biggr.\\
&\quad\times\biggl.\left(\omega_{f_0}+\omega_D+\omega_{P_1}\right)-\omega_{P_1}\left(k^0\right)^2\biggr],
\end{align}

\begin{align}\nonumber
\mathcal{N}_1=&\omega_{P_1}\Biggl[P^0\omega_{D} \left(\omega_D+\omega_{P_1}\right)\left(2\omega_{f_0}+\omega_D+\omega_{P_1}\right)+\left(P^0\right)^2\omega_{f_0}k^0-P^0\left(2\omega_{f_0}+\omega_D\right)\left(k^0\right)^2\Biggr.\\
-&\left(\omega_{f_0}+\omega_D\right)k^0\left(\omega_{f_0}^2+\left(\omega_{P_1}+\omega_D\right)^2+\omega_{f_0}\left(2\omega_{P_1}+\omega_{D}\right)-\left(k^0\right)^2\right)\Biggl. \Biggr],
\end{align}
and 
\begin{align}\nonumber
\mathcal{D}=&2\omega_{P_1}\omega_{f_0}\omega_{D}\left(P^0-\omega_{f_0}-\omega_{D}+i\epsilon\right)\left(-P^0-\omega_{f_0}-\omega_{P_1}+k^0+i\epsilon\right)\left(-\omega_{P_1}-\omega_{D}+k^0+i\epsilon\right)\\
\times&\left(P^0-\omega_{f_0}-\omega_{P_1}-k^0+i\epsilon\right)\left(\omega_{P_1}+\omega_{D}+k^0\right)\left(P^0+\omega_{f_0}+\omega_{D}\right).\label{eqdeno}
\end{align}
Eventually, in the calculations, we replace $-\omega_{f_0}+i\epsilon \to -\omega_{f_0}+i\Gamma_{f_0}/2$ in Eq.~(\ref{eqdeno}), to take into account the unstable nature of $f_0$.

To summarize, we calculate the amplitude in Fig.~\ref{diagram1}a as
\begin{align}\nonumber
t_a=&
2i ~g_{D(2900)^0\to D^0f_0}~  g_{f^0\to P_1P_2}~ \mathcal{I}_{VPP}~g_{VPP}~\frac{\epsilon_\mu \left(k\right)P^\mu}{\left(k^0 P^0\right)^2-k^2 P^2} \Biggl\{P^0 k^0~\biggl[k^0\int\frac{d^3q}{\left(2\pi\right)^3}\left(-i\frac{\mathcal N_1}{\mathcal D}\right)\Biggr.\biggr.\\
-&\Biggl.\biggl.\int\frac{d^3q}{\left(2\pi\right)^3}\left(\vec k\cdot \vec q\right)\left(-i\frac{\mathcal N_0}{\mathcal D}\right)\biggr]-k^2~P^0\int\frac{d^3q}{\left(2\pi\right)^3}\left(-i\frac{\mathcal N_1}{\mathcal D}\right)\Biggr\}.
\end{align}

The amplitude for the decay process shown in Fig.~\ref{diagram1}b can be written by replacing the VPP vertex by the coupling $g_{D_{s0}^*(2317)\to D K}$ in Eq.~(\ref{lagr_a}), finding
\begin{eqnarray}\nonumber
t_b&=&
i ~g_{D(2900)^0\to D^0f_0}~  g_{f^0\to P_1P_2}~g_{D_{s0}^*(2317)\to D K}  \Biggl\{\int\frac{d^4q}{\left(2\pi\right)^4} \frac{1}{\left[\left(k+q\right)^2-m_{D}^2\right]} \Biggr.\\\nonumber
&\times&  \frac{1}{\left[\left(P-k-q\right)^2-m_{f_0}^2\right]\left[q^2-m_{P_1}^2\right]}\Biggr\}\\
&=&i ~g_{D(2900)^0\to D^0f_0}~  g_{f^0\to P_1P_2}~g_{D_{s0}^*(2317)\to D K}\int\frac{d^3q}{\left(2\pi\right)^3}\left(-i\frac{\mathcal N_0}{\mathcal D}\right).
\end{eqnarray}

\section{Results and discussions}
Having calculated the amplitudes, we can determine the partial decay widths of $D(2900)$ using Eq.~(\ref{eq:width}). Before showing the results, we must discuss the  uncertainties present in the formalism.   Among the couplings given in Table.~\ref{couplings},  besides taking the uncertainty on the value for $D_{s0}^*(2317)^+\to D^{0}K^{+}$  from Ref.~\cite{Gamermann:2006nm}, we consider a 10$\%$ error on the other couplings too. Such an error on  the $D(2900)^0\to D^0f_0(980)$ coupling is consistent with varying the width of $D(2900)$ in 55~$\pm$~10  MeV.  Additionally, we take the mass for $D(2900)$ in the range 2900~$\pm$~50 MeV and for $f_0(980)$ as 990~$\pm$~20 MeV~\cite{pdg}. To take into account all the uncertainties,  random numbers are generated within the range of all the inputs and  mean values as well as standard deviations on  the results are evaluated.

The results obtained are given in Table~\ref{results}. 
\begin{table}[h!]
\caption{Partial widths of $D(2900)$ to the main two-body decay channels. }\label{results}
 \begin{tabular}{ccc}
 \hline
Decay channel&Decay width (MeV)\\
\hline\hline
$D^{*0}\pi^0$& $0.18\pm0.04$\\
$D^{*+}\pi^-$& $0.35\pm0.07$\\
$D_s^{*+}K^-$&$0.44\pm0.10$\\
$D^*_{s0}(2317)^+ K^-$&$18.33\pm7.25$\\
\hline
\end{tabular}
\end{table}
It can be seen that the decay width to a $D^*_{s0}(2317)^+ K^-$ final state is the largest of all, it turns out to be about 40-100 times bigger than the widths to the other channels. Such findings imply that $D^*_{s0}(2317)^+ K^-$, rather than $D^*\pi$ analyzed in Ref.~\cite{Aaij:2013sza}, should be a far more promising channel to look for a signal of $D(2900)$ which is a $Df_0(980)$ moleculelike state.  

\begin{figure}[h!]
\includegraphics[width=0.75\textwidth]{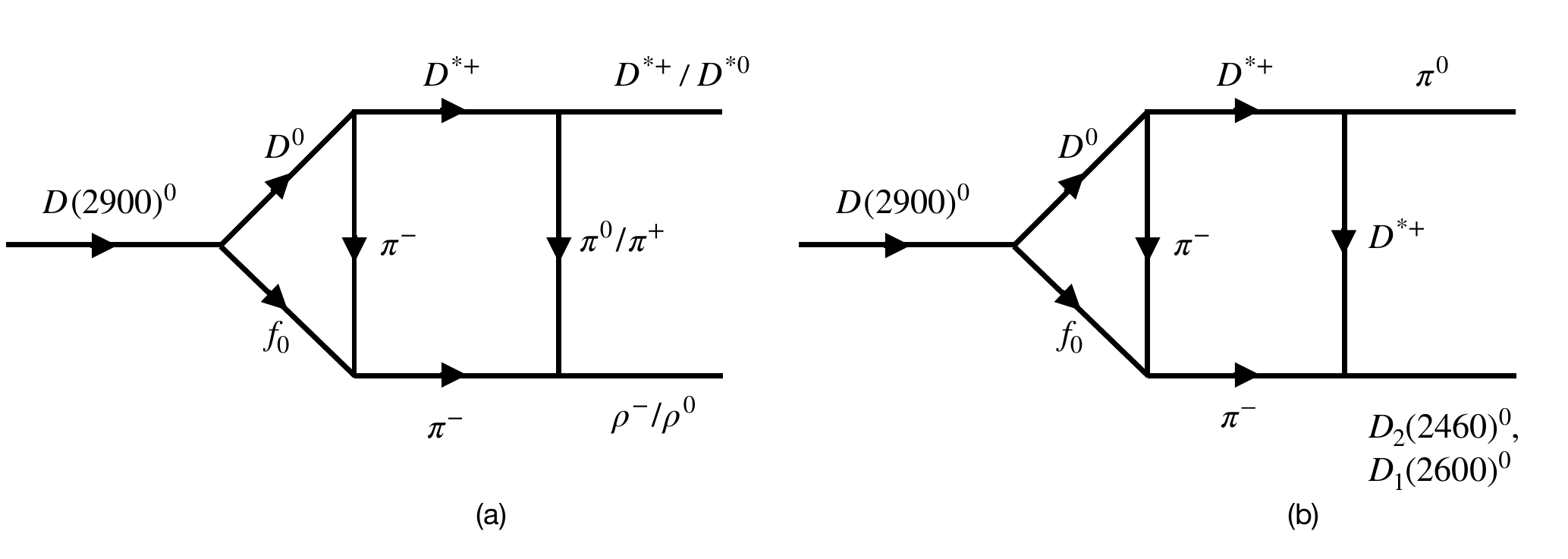}
\caption{Decay mechanism of $D(2900)^0$ to final states like $D^{*+}\rho^-$, $D^{*+}\rho^-$, $D^{*0}_2(2460)\pi^0$ and $D^{*0}(2600)\pi^0$.}\label{diagram3}
\end{figure}

We would now like to discuss that the mechanisms of  decay of $D(2900)$ to  final states like $D^*\rho$, $D\rho$, $D^*_2(2460) \pi$, $D_1^*(2600) \pi$ involve higher order loops, due to the $Df_0(980)$ molecular nature of $D(2900)$.  We show some examples in Fig.~\ref{diagram3}
of the  decay processes to the mentioned final states. Similar will be the mechanisms to yet other channels, like $D^*\omega$, $D\omega$. Such mechanisms imply a suppressed partial widths to such channels. Thus, our state can be distinguished from the states predicted within the quark model calculations~\cite{Ebert:2009ua,Sun:2013qca,Yu:2014dda,Lu:2014zua,Xiao:2014ura,Godfrey:2015dva,Song:2015fha,Batra:2015cua,Li:2017zng,Gupta:2018zlg,Badalian:2020ngz,Gandhi:2021col}.  We hope that our present study can be useful in investigation of charm meson in the region around 3000 MeV.

\section*{Acknowledgements}
B.B. M.,  K.P.K and A.M.T gratefully acknowledge the  support from the Funda\c c\~ao de Amparo \`a Pesquisa do Estado de S\~ao Paulo (FAPESP), processos n${}^\circ$ BRENDA's CONTRACT NUMBER      2019/17149-3 and 2019/16924-3. K.P.K and A.M.T are also thankful to the Conselho Nacional de Desenvolvimento Cient\'ifico e Tecnol\'ogico (CNPq) for grants n${}^\circ$ 305526/2019-7 and 303945/2019-2.

\end{document}